\documentclass[%
 twocolumn,
superscriptaddress,
 amsmath,amssymb,
 aps,
floatfix,
longbibliography,
]{revtex4}

\usepackage{graphicx}
\usepackage{dcolumn}
\usepackage{bm}
\usepackage{nowidow}
\usepackage{tabularx}
\usepackage{siunitx}
\usepackage[dvipsnames]{xcolor}
\usepackage[colorlinks=true, allcolors=black]{hyperref}

\usepackage{xcolor}

\usepackage{natbib}
\setcitestyle{square, comma, numbers,sort&compress, super}
\begin{document}

\newcommand{\mum}[1]{${#1}$\,\textmu m} 

\title{Polycrystalinity enhances stress build-up around ice}

\author{Dominic Gerber}
\affiliation{Department of Materials, ETH Z\"{u}rich, 8093 Z\"{u}rich, Switzerland.}%

\author{Lawrence A. Wilen}
\affiliation{Center for Engineering Innovation and Design, School of Engineering and Applied Sciences, Yale University, New Haven, Connecticut 06520, USA}%

\author{Eric R. Dufresne}
\affiliation{Department of Materials, ETH Z\"{u}rich, 8093 Z\"{u}rich, Switzerland.}%

\author{Robert W. Style}
\affiliation{Department of Materials, ETH Z\"{u}rich, 8093 Z\"{u}rich, Switzerland.}%

\begin{abstract}
Damage caused by freezing wet, porous materials is a widespread problem, but is hard to predict or control.
Here, we show that polycrystallinity makes a great difference to the stress build-up process that underpins this damage. 
Unfrozen water in grain-boundary grooves feeds ice growth at temperatures below the freezing temperature, leading to the fast build-up of localized stresses.
The process is very variable, which we ascribe to local differences in ice-grain orientation, and to the surprising mobility of many grooves -- which further accelerates stress build-up.
Our work will help understand how freezing damage occurs, and in developing accurate models and effective damage-mitigation strategies.
\end{abstract}

\maketitle

Damage caused by ice growth can arise by two distinct mechanisms. 
When water freezes in a closed system, like a full bottle of water, damage occurs due to the $\sim$9\,\% volumetric expansion.
This expansion pushes on the water's surroundings, causing the pressure to rise.
While the water remains partially unfrozen, this pressure will continue to increase by $\sim$11\,MPa per degree of undercooling~\cite{style_generalized_2023}.
This mechanism is almost unique to ice: most other liquids shrink as they freeze, reducing their pressure until the liquid cavitates~\cite{kohler_cavity_2018, scoti_crystallization_2022}.
When water instead freezes in an open system, such as in a pore in the soil, damage can occur through the lesser-known process of \emph{cryosuction} \cite{unold_cryosuction_2018, derjaguin_flow_1986, wettlaufer_premelting_2006,ozawa_segregated_1989, coussy_mechanics_2010}.
Now, water can flow away as ice grows, preventing any pressure build-up during ice's initial growth.
However, if the ice is below its freezing temperature, and in contact with an unfrozen supply of water, it will subsequently suck water back into the pore, causing the ice to grow~\cite{gerber_stress_2022, wilen_frost_1995, wettlaufer_premelting_2006}.
Then, the ice can push open the pore, causing pressure to build up with a maximum pressure of about 1\,MPa per degree of undercooling~\cite{gerber_stress_2022,style_generalized_2023, wang_frost-heaving_2018, wettlaufer_ice_2013, black_applications_1995, derjaguin_flow_1986, kjelstrup_transport_2021}.
Importantly, the expansion of the pore due to cryosuction can theoretically be unbounded, provided enough unfrozen water is available.
Cryosuction is responsible for much of the damage caused by freezing wet, porous solids~\cite{wang_frost-heaving_2018, wettlaufer_ice_2013,carter_soil_2016, coussy_mechanics_2010, dash_physics_2006}, and occurs in any liquid~\cite{taber_mechanics_1930, hiroi_frost-heave_1989, dash_frost_1991}.

Although the basic mechanisms underlying freezing-induced stresses are well understood, it is challenging to reliably predict how and where these appear.
For example, for freezing of a particular soil type, it is not possible to predict how fast, and in what form ice will grow \cite{dimillio_quarter_1999, dash_physics_2006,peppin_physics_2013, carter_soil_2016,style_kinetics_2012, oneill_physics_1983}.
This is despite the availability of a wide range of frost-heave models (e.g. \cite{schollick_segregated_2016,you_situ_2018,watanabe_amount_2002,zhou_ice_2020,vignes_model_1974, gilpin_model_1980, derjaguin_flow_1986, rempel_microscopic_2011, peppin_physics_2013,style_kinetics_2012,konrad_mechanistic_1980,kjelstrup_transport_2021,wettlaufer_theory_1996, wettlaufer_premelting_2006,vlahou_freeze_2015,wettlaufer_dynamics_1995,gagliardi_nonequilibrium_2019, oneill_physics_1983, dash_frost_1991, wang_frost-heaving_2018, everett_thermodynamics_1961a}).
As a result, our understanding of freezing is almost exclusively empirical -- in fields ranging from civil engineering and road design, to cryopreservation, agriculture, food science, medicine, and low-temperature biology~\cite{carter_soil_2016, 
dimillio_quarter_1999, 
benson_cryopreservation_2008, 
deville_freezing_2017-1}.
This suggests some aspects of the freezing process are not fully understood.

Here, we show that ice polycrystallinity, a typically overlooked factor, can play a dramatic role by accelerating the build-up and expanse of freezing-induced stresses.
Water-filled grain-boundary grooves in ice act as conduits that feed ice growth across the surface of polycrystalline ice.
This process results in the formation of large, highly localized stresses which can lead to damage.
The dynamics of the stress build-up can vary greatly between grooves but the resulting stresses are always larger than stresses that appear around monocrystalline ice.
Furthermore, we observe that many grooves are mobile (often in an unpredictable manner), and these grooves support even faster ice growth, with greater potential for damage.

\begin{figure}[!htb]
    \centering
    \includegraphics[width=1\linewidth]{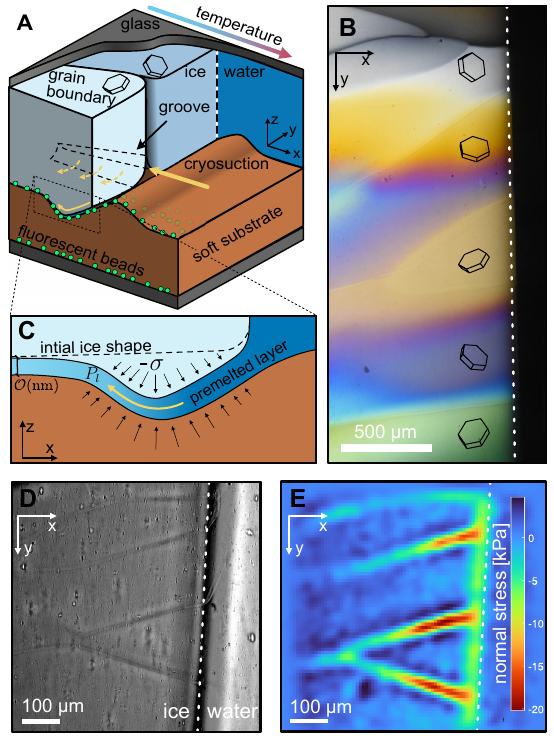} 
    \caption{
    (\textbf{A}) Schematic of the experimental cell containing polycrystalline ice
    (\textbf{B}) An ice-water interface imaged through crossed-polarizers, highlighting the different crystal orientations of the individual grains. Crystal symbols are for illustrative purposes. 
    (\textbf{C}) Schematic of the cryosuction process into a premelted layer.
     (\textbf{D}) Grain boundaries can be clearly seen in an enhanced-contrast, bright-field micrograph of ice in the cell (thin lines). 
     (\textbf{E}) Stresses below the ice in (D). The image is taken 10~min after the start of the experiment. Substrate stiffness:~265\,kPa, substrate thickness:~\mum{130}, ice thickness:~\mum{500}, temperature gradient:~0.4\,K/mm.}
    \label{fig:stress_under_groove}
\end{figure}

We study the role of polycrystallinity in freezing damage by growing ice in the set-up shown in Fig.~\ref{fig:stress_under_groove}A~\cite{gerber_stress_2022}.
This consists of an open-ended, water-filled Hele-Shaw cell with a lower surface coated in a soft, silicone layer.
The cell is placed in a temperature gradient, so that ice fills the cell's left-hand side, when viewed in the $x$,$y$ plane (see Fig.~\ref{fig:stress_under_groove}).
Any stresses that develop around the ice can be observed as deformations to the silicone layer.
We measure these over time with a confocal microscope by imaging the 3-D positions of fluorescent nanoparticles that are attached to the top and bottom of the silicone layer~\cite{xu_imaging_2010}.
This allows us to calculate both displacement maps of the silicone surface, and corresponding maps showing the stress build-up around the ice.
Stresses are calculated from displacements via traction force microscopy (TFM), essentially by solving an elasticity problem (see Supplement, \cite{style_traction_2014, gerber_stress_2022}).
Stress and displacement maps show similar qualitative features (see Fig.~S2 in the Supplement), but stress maps have lower resolution due to smoothing in the TFM algorithm.
Therefore, in this paper, we predominantly present displacement data.

Ice that forms in the cell is naturally polycrystalline.
Ice grains are randomly oriented, and have an average size of $O$(\mum{100}), comparable to the ice thickness~\cite{mullins_effect_1958}.
Individual birefringent, ice grains appear with different colors in crossed polarizers (Fig.~\ref{fig:stress_under_groove}B)~\cite{zhang_single_2021, gerber_stress_2022, sorensen_revised_2013}.
At the same time, water appears black -- the bulk ice-water boundary is shown as the dashed white line in Fig.~\ref{fig:stress_under_groove}.
Grain boundaries can also be seen without polarizers, as they appear as darker lines in images (Fig.~\ref{fig:stress_under_groove}D).

We perform experiments by growing ice up to the middle of the cell, and then holding it fixed in a constant temperature gradient.
When ice appears, it initially exerts only minor stresses on its surroundings (see Supplement Fig. S3~\cite{gerber_stress_2022}), but subsequently these stresses grow steadily.
Fig.~\ref{fig:stress_under_groove}E shows the normal stresses exerted by the ice on the underlying substrate, 10~minutes after initial ice formation.
We see a small stress build-up just on the cold side of the ice-water interface -- a vertical green band in the data.
This type of `ice-front' stress build-up has been previously reported~\cite{gerber_stress_2022, wilen_frost_1995, dash_premelting_1995}.
However, these stresses are dwarfed by the stresses under the grain boundaries, which at $O(20\,$kPa$)$, and being highly localized, are easily capable of breaking many soft materials.
The grain-boundary stresses also extend back to cold temperatures, allowing stresses to build up across a broad area of ice-substrate interface.

Stress development is caused by the localized accumulation of ice, fed by unfrozen water at the ice-substrate.
This water appears in two forms.
Firstly, nanometric, \emph{premelted layers} of water exist between ice and a neighboring substrate -- shown schematically in Fig.~\ref{fig:stress_under_groove}C~\cite{wettlaufer_premelting_2006, wilen_dispersion-force_1995-1, slater_surface_2019, dash_frost_1991, churaev_disjoining_1994}.
There, interfacial forces combine to give a repulsive interaction between ice and substrate, which act to drag in liquid from nearby sources of bulk water.
This mechanism underlies the `ice-front' stress band in Fig.~\ref{fig:stress_under_groove}E.
These premelted films thicken into macroscopic `grain-boundary grooves' at the triple junction where two ice grains meet at a substrate (see Fig.~\ref{fig:stress_under_groove}A).
Grooves remain unfrozen below 0\,°C because of the surface energy of the ice-water interface, and are analogous to Plateau borders in foams \cite{style_surface_2005, wilen_giant_1995, mullins_theory_1957}.
Groove width is predicted to decrease inversely proportional to the local undercooling \cite{wilen_giant_1995}.
In our experiments, grooves are always much larger than premelted films: 100\,nm radius tracer particles are easily transported along them (Supplementary Video V1).

\begin{figure}
    \centering
    \includegraphics[width=1\linewidth]{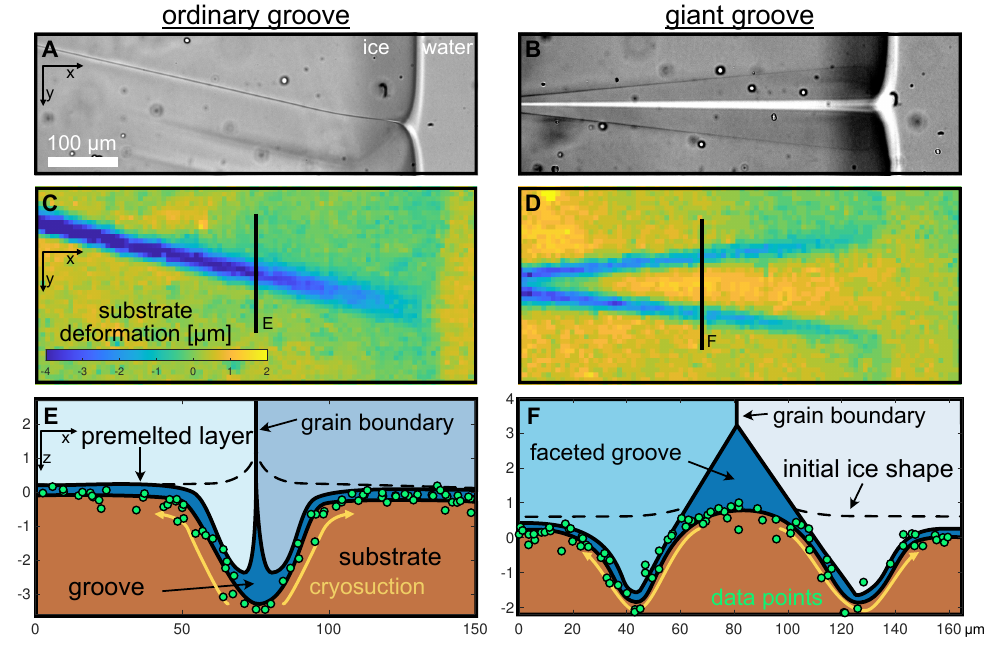} 
    \caption{
    (\textbf{A}, \textbf{B}) Bright-field images of ordinary and giant grooves.
    (\textbf{C}, \textbf{D}) substrate indentations caused by ice growth.
    (\textbf{E}, \textbf{F}) cross-sections through the data at the lines indicated in (C,D), with schematics of the corresponding grain-boundary grooves. Images are 30\,mins after the start of the experiment. Substrate stiffness: 38\,kPa, substrate thickness~\mum{100}, ice thickness:~\mum{200},  temperature gradient: 0.1\,K/mm. }
    \label{fig:groove_schematic}
\end{figure}

Stress development at grain-boundary grooves appears to be driven by ice growth at their sides.
This is clear when we observe `giant', faceted grooves.
In these rare grooves, adjoining ice grains expose their basal facet towards the groove.
The basal facets only grow at a larger undercooling of about 0.03\,°C, yielding grooves that are much larger than ordinary grooves~\cite{wilen_giant_1995, dash_dynamics_1999, libbrecht_physical_2017}.
Fig.~\ref{fig:groove_schematic} shows examples of both ordinary and giant grooves, along with the associated substrate indentations.
The indentation under the ordinary groove appears as a single trough.
However, under the giant groove, the ice only grows into the substrate at the groove edges.
In the middle of the giant groove, the substrate actually bulges back upwards -- as shown in Fig.~\ref{fig:groove_schematic}F, with a schematic cross-section through a groove (green dots show surface displacement data).
This indicates that stresses only build up at the sides of the groove, where the ice and substrate come into close contact.
Here, there is a premelted layer, which can drag in water to feed ice growth via the same mechanism that underlies `ice-front' stress build-up (compare Fig.~\ref{fig:stress_under_groove}C with Figs.~\ref{fig:groove_schematic}E,F).
We expect that something similar is occurring at ordinary grooves (Fig.~\ref{fig:groove_schematic}E), but that the scale is too small to resolve individual bulge and troughs.

\begin{figure}
    \centering
    \includegraphics[width=1\linewidth]{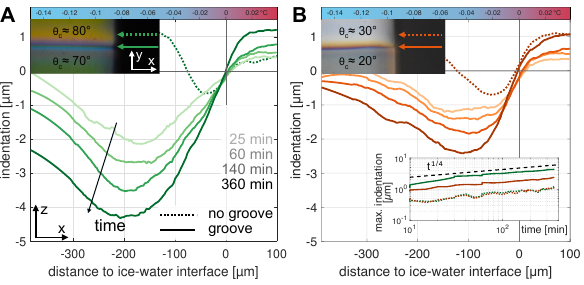} 
    \caption{Evolution of the indentation directly under (continuous curves) and parallel to (dashed curves) two grain-boundary grooves (\textbf{A},\textbf{B} respectively) under identical experimental conditions. 
    (Insets top left): two images, taken using crossed polarizers, showing the grain boundaries. $\theta_c$ is the angle between the c-axis of the ice crystal and the substrate, estimated from the birefringent color of the grains \cite{gerber_stress_2022}. 
    (Inset in \textbf{B}) Time evolution of the maximum indentation. 
    Substrate stiffness:~280\,kPa, substrate thickness:~\mum{100}, ice thickness:~\mum{500},  temperature gradient: 0.4\,K/mm. }
    \label{fig:groove_evolution}
\end{figure}

Indentations under the ice continuously grow, while the extent of the indentated area gradually expands (Supplementary Video V2).
For example, Fig.~\ref{fig:groove_evolution}A shows the evolution of surface indentations under the center-line of a stationary grain boundary.
For comparison, the much smaller indentations that form away from the grain boundaries are also shown at the last time-point (dotted curves, t=~360~min).
All curves exhibit a maximum indentation away from the bulk ice-water interface ($x=0$), and this continuously grows, and moves back to colder temperatures.
Close to the ice-water interface, the indentation appears to stall at a value that is proportional to the undercooling (i.e. the distance from the interface).

While the qualitative features of ice growth are very repeatable, the dynamics can vary greatly between grooves.
Fig.~\ref{fig:groove_evolution}B shows a stationary groove near the one in Fig.~\ref{fig:groove_evolution}A, but between crystals of different orientation.
The indentation below this groove is much smaller.
The inset in Fig.~\ref{fig:groove_evolution} shows the maximum indentation under the two grooves (continuous curves).
In this case, the two grooves show approximately power-law indentation growth, with exponents $\sim 0.25$, but with very different magnitudes.
Interestingly, the indentations far from grooves also follow this power-law (dashed curves).
However, this exponent is not universal, as in other experiments, exponents range from $0.25$ to $0.33$ (see Supplement Fig.~S4).

\begin{figure}
    \centering
    \includegraphics[width=1\linewidth]{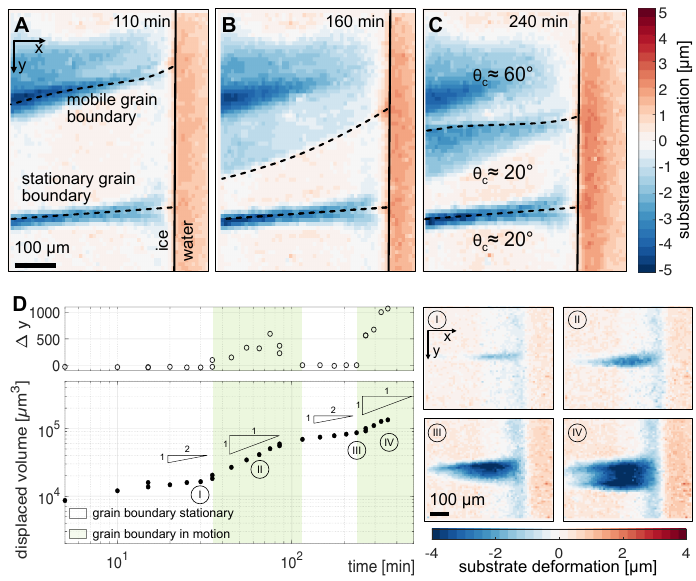} 
    \caption{
    (\textbf{A}-\textbf{C}) Substrate indentations below a body of ice with a stationary (top) and a mobile (bottom) grain boundary moving in a stop-start fashion (Supplementary Video V2). Dashed curves show the current location of the grain boundaries.
    Ice thickness: $\sim$\mum{200}, substrate stiffness: 52\,kPa, substrate thickness: \mum{45}, temperature gradient: 0.1\,K/mm.
    (\textbf{D}) Evolution of the displaced volume below a sporadically moving groove (green sections indicate groove motion).
    Top panel: the $y$-position of the grain boundary close to the bulk ice-water interface.
    Panels I-IV show surface displacements at the corresponding time points. 
    Ice thickness: $\sim$\mum{500}, substrate stiffness: 52\,kPa, substrate thickness: \mum{45}, temperature gradient: 0.1\,K/mm.
    } 
    \label{fig:groove_movement}
\end{figure}

Roughly half of all grain boundaries at the ice-water interface are also mobile, and this can significantly accelerate stress accumulation.
We observe that grain boundaries can translate steadily in one direction, move with stop-start motion, oscillate about a fixed position, oscillate with intermittent pauses, exhibit unpredictable combinations of all the above, or fuse with other grain boundaries (e.g. Fig.~\ref{fig:groove_movement}A-C, Supplementary Videos V2-5).
This is important, as ice growth rates are enhanced under mobile grain boundaries.
For example, Fig. \ref{fig:groove_movement}D shows the total accumulated volume of ice, $V$, under a sporadically moving grain boundary.
$V$, increases significantly faster in mobile phases ($V\sim t$), than in stationary phases ($V\sim t^{1/2}$).
Ice growth always occurs directly at grain boundaries, and does not melt back if the grain boundary moves on.
Thus, mobile grain boundaries result in widely distributed stresses as (Fig. \ref{fig:groove_movement}A-C, Supplementary Videos V2, V3), and have a very different potential for damage than stationary grooves.

Grain-boundary motion appears to be linked to grain orientation and to the presence of a soft substrate.
Certainly some motion is driven by grain coarsening that reduces the interfacial energy within the ice~\cite{hondoh_anisotopy_1979, knight_grain_1966, nasello_temperature_2005, higashi_structure_1978, mullins_effect_1958, verma_computational_2022, chen_grain_2020,min_effect_2006}, but this does not explain motion that reverses direction.
In general, we observe that boundaries between grains with large differences in crystallographic orientation were more mobile than boundaries between grains with similar orientations (\emph{e.g.}~\cite{min_effect_2006}).
Furthermore, oscillatory motion was only seen on soft substrates (e.g. Supplementary Video V5), and sudden jumps in grain-boundary position were often accompanied by abrupt changes in the substrate deformation (Supplementary Video V3).
This suggests the presence of substrate-mediated instabilities of the grain-boundary position.
However, we leave exploration of this to future work.

We can qualitatively explain many of the features of our observations using existing theory for cryosuction in premelted films~\cite{style_generalized_2023, zhou_ice_2020, wettlaufer_premelting_2006, derjaguin_flow_1986, everett_thermodynamics_1961a, gilpin_model_1980}.
This suggests that ice will continue to grow by suction into films, until it reaches a temperature-dependent stress given by the Clapeyron equation (which describes thermodynamic equilibrium between ice and water):
\begin{equation}
    -\sigma - P_a=\rho q_m\frac{T_m-T}{T_m}.
   \label{eqn:clausius}
\end{equation}
Here, $\rho$ and $q_m$ are the density and latent heat of melting of ice, $T_m$ is the bulk freezing temperature at atmospheric pressure, $P_a$ is the pressure of the nearby source of bulk water (here atmospheric pressure), and $\sigma$ is the normal stress exerted by the ice on the substrate ($\sigma<0$ when compressing the substrate)~\cite{gerber_stress_2022, style_generalized_2023}.
This shows that ice can push harder on its surroundings at colder temperatures, and hence we observe larger stresses at the undercooled grain boundaries than we observe near the bulk ice-water interface (e.g. Fig. 1E).

Although the Clapeyron equation tells us the maximum stress that can occur in freezing systems, ultimately stress build-up is governed by the dynamics of water transport.
Water cannot travel large distances along nanoscopic, premelted films, due to viscous drag \cite{wettlaufer_dynamics_1995, wilen_frost_1995, derjaguin_flow_1986, wettlaufer_premelting_2006}.
Thus, stresses can only build up near easy supplies of bulk water -- i.e. either close to the bulk ice-water interface or adjacent to grain-boundary grooves.
Hence, the novel function of the grooves is to transport water to cold temperatures well behind the bulk ice-water interface, greatly expanding the spatial extent of stress build-up.    
At colder temperatures, the mobility of water in the premelted layer will be hugely reduced, as the film thickness, $h$, drops off dramatically with temperature \cite{slater_surface_2019}, and flow rates $\sim h^3$ \cite{style_surface_2005}.
This explains why we see slower ice growth at colder temperatures (e.g. Fig. \ref{fig:groove_evolution}).
Ultimately, we expect premelted film thickness to vanish at a certain undercooling, at which point all stress accumulation should disappear.
Interestingly, premelted film thickness at a substrate is known to depend strongly on the local ice grain orientation \cite{slater_surface_2019, dosch_glancing-angle_1995,furukawa_ellipsometric_1987}.
This would lead to large differences in the dynamics of stress build-up between differently orientated grains, and can potentially explain the large variability we see in our experiments.
In the future, carefully observing this orientation dependence may offer a way to measure premelted film thicknesses as a function of temperature and grain orientation, and to see when films vanish -- properties which proven elusive to characterization~\cite{slater_surface_2019, chen_carefull_2019}, but which are important for applications ranging from ice rheology to cryobiology and food science~\cite{persson_ice_2015, mohlmann_are_2009, dash_premelting_1995}.

In conclusion, we have shown that freezing-induced stress build-up is dramatically enhanced in polycrystalline ice relative to monocrystalline ice.
Grain-boundary grooves act as conduits for unfrozen water that feed ice growth across ice-substrate interfaces.
This growth is very localized to the grooves, and can quickly reach pressures of several 10s of kPa: easily large enough to break many materials.
Pressures should stall at a value proportional to local undercooling, so near deeply undercooled grooves, there will be a high damage potential.
Thus, this process can play a key role in freezing stress development, and should be accounted for in models.
Interestingly, we see large variability in how fast stress builds up, and this may explain difficulties in producing accurate models of freezing.
We ascribe this stochasticity to differences in ice-grain orientation, and to the mobility of some grain boundaries.
A key question for future work is if there is a predictable average behavior, to allow accurate incorporation into theoretical models.

Our results have important consequences for understanding damage in freezing materials.
For example, the pressure distribution applied by ice to a confining material will determine how this material breaks.
Thus, especially in brittle materials, stress localization at grooves may play a key role in determining when and how damage occurs.
Additionally, we expect ice's growth history to effect stress build-up.
Fine-grained ice will have more grooves where stress can develop than coarser-grained ice, and thus is likely to exert more cumulative forces on its surroundings.
Ice nucleated at large undercoolings will tend to form many small grains, while slowly-grown ice will have fewer, large grains.
Thus, altering how ice first forms and then ages could alter how stress builds up.
Additionally, the stress build-up process should be significantly changed in the presence of chemicals that bind to different ice facets and inhibit recrystallization (e.g. antifreeze proteins~\cite{tas_freezer_2021, bayer-giraldi_growth_2018-1, bar_dolev_ice-binding_2016, knight_adsorption_2001}).
Finally, we anticipate that the ice-growth process presented here will also occur at the water-filled ice veins at triple junctions between ice grains inside the ice bulk.
Cryosuction should also suck water into the grain boundaries adjacent to these veins.
This would effectively cause a volumetric expansion of the ice, driving even faster stress accumulation around the ice.

We thank Christine McCarthy and Charlotta Lorenz for helpful discussions. This work was supported by an ETH Research Grant (ETH-38~18-2), and the Swiss National Science Foundation (200021-212066).

%

\end{document}